\documentclass[referee]{raa}            

\usepackage{graphicx,times}             
\usepackage{natbib}
\usepackage{amssymb,amsmath}
\bibpunct{(}{)}{;}{a}{}{,}

\usepackage[a4paper=true,dvipdfm=true,pagebackref=true]{hyperref}
\hypersetup{colorlinks = true, linkcolor = green, anchorcolor = red, citecolor = blue, filecolor = red, pagecolor = red, urlcolor = red}

\begin{document}

   \title{The application of co-integration theory in ensemble pulsar timescale algorithm
}

   \volnopage{Vol.0 (20xx) No.0, 000--000}      
   \setcounter{page}{1}          

  \author{Feng Gao
      \inst{1,2,3,4}
   \and Ming-Lei Tong
      \inst{1,2}
   \and Yu-Ping Gao
      \inst{1,2,3}
   \and Ting-Gao Yang
      \inst{1,2}
   \and Cheng-Shi Zhao
      \inst{1,2}
   }

    \institute{National Time Service Center, Academy of Sciences, Xi'an 710600, China; {\it fengg@xust.edu.cn; \it mltong@ntsc.ac.cn}\\
        \and
             Key Laboratory of Time and Frequency Primary Standards, Chinese Academy of Sciences, Xi'an  710600, China\\
        \and
             University of Chinese Academy of Sciences, Beijing 100049, China\\
        \and
            Department of Applied Physics, Xi'an University of Science and Technology, Xi'an  710054, China\\
\vs\no
   {\small Received~~20xx month day; accepted~~20xx~~month day}}

\abstract{ Employing multiple pulsars and using an appropriate algorithm to establish ensemble pulsar timescale  can
reduce the influences of various noises on the long-term stability of pulsar timescale, compared to a single pulsar.
However, due to the low timing precision and the significant red noises of some pulsars, their participation in the
construction of ensemble pulsar timescale is often limited. Inspired by  the principle of solving non-stationary
sequence modeling using co-integration theory, we puts forward an  algorithm based on the co-integration theory
to establish ensemble pulsar timescale. It is found that this algorithm can effectively suppress some noise sources
if a co-integration relationship between different pulsar data exist. Different from  the classical  weighted average
algorithm, the co-integration method provides the chances of the pulsar with significant red noises to attend the
establishment of ensemble pulsar timescale.  Based on the   data from the North American Nanohertz Observatory for
Gravitational Waves, we found that  the co-integration algorithm  can successfully reduce  several timing noises and
improve the long-term stability of the ensemble pulsar timescale.
}

   \authorrunning{F. Gao,M.-L. Tong, \& Y.-P. Gao, et al. }            
   \titlerunning{Ensemble Pulsar Timescale Algorithm }  

   \maketitle

%
%
\section{Introduction}           
\label{sect:intro}
Pulsar timing  is an effective  tool in studying  astrophysics and fundamental physics.
These include tests of gravitation, precision constraints of general relativity, and
especially using arrays of pulsars as detectors of low-frequency gravitational wave
(\citealt{Zhu+etal+2015,Will+2014,Arzoumanian+etal+2015a}). The basis of  pulsar timing
is the high-precision timing model, accomplished by the determination of a series model
parameters, such as the spin parameters, astrometric parameters, binary orbit parameters
and so on. The errors of the  model parameters will affect the timing precision in different
ways(\citealt{Tong+etal+2017}). At present,  millisecond pulsars (MSPs) have higher stability
of rotation and are more widely used in the study of pulsar time scale
(\citealt{Splaver+2004,Verbiest+etal+2009}). For example, G. Hobbs (\citealt{Hobbs+etal+2012})
obtained a preliminary pulsar time scale based on Parkes Pulsar Timing Array including 19
millisecond pulsars observed by Parkes radio telescope. It was shown that pulsar timing
array allows investigation of ``global'' phenomena, such as a background of gravitational
waves or instabilities in atomic time scales that produce correlated timing residuals in
the pulsars of the array. However, there are various physical processes that might be
responsible for the accuracy of pulsar time scale, timing noise is still not fully
understood, but usually refers to unexplained low-frequency features in the timing
residuals of pulsars. In the presence of red timing noise, W. Coles (\citealt{Coles+etal+2011})
adopted a transformation based on the Cholesky decomposition of the covariance matrix that
whitens both the residuals and the timing model, which has sufficient accuracy to optimize
the pulsar timing analysis. In addition, using data from multiple pulsars, it is possible
to obtain an average pulsar time scale that has a stability better than those
derived from individual pulsar data(\citealt{Petit+etal+1993,Rodin+2008,Zhong+Yang+2007,Hobbs+etal+2010}).

The purpose of using multiple pulsars data is to suppress  the timing noise intensity of
individual pulsar. It will open up a new window to improve the accuracy and long-term
stability of pulsar timescale by establishing ensemble pulsar timescale (EPT). However, similar to the establishment of atomic time scale (AT)
the accuracy and long-term stability of EPT depend significantly on the information of
timing residuals of pulsars data involved. The lower the timing noise is, the better
the result of EPT will be obtained, which often limits the participation of a large number
of pulsars with significant timing noise. Based on the timing clock model analysis, any
clock model can be regarded as the connection between the regression model and  time
series variables. Whether establishing a well pulsar timescale or make its forecast, the clock
difference series should be stationary. Only in this way can we ensure that some statistic
parameters in the selection and examination of the model, such as determinable coefficient
$R^2$ and $T$ statistics, have standard normal distribution, and therefore, the statistics
are reliable. Otherwise, all of the above inferences can easily produce a spurious regression.

In the economics fields, in order to avoid spurious regression of non-stationary series,
Engle and Granger proposed the co-integration theory that provided another way for the
modeling of non-stationary series(\citealt{Engle+Granger+1987}). For example, although
some economic variables themselves are non-stationary series, but linear combination
of them is likely to be stationary. This combining process is known as co-integration equation,
and it can explain the long-term equilibrium relationship between different variables.
In principle, the pulsars with significant timing noise show the non-stationary characteristics
of timing residuals, if the linear combination of pulsars timing residuals is stationary
series, they are also co-integration. According to the above ideas, an algorithm based on
co-integration theory to establish EPT is proposed in this paper, which mainly use the
pulsars with significant timing noise, and the results show that the algorithm can
successfully reduce several timing noises and improve significantly the long-term frequency
stability of EPT. Co-integration is a powerful method, because it not only allows us to
characterize the equilibrium relationship between two or more no-stationary series, but
also will provide a better guidance in studying the establishment of  EPT in future.

\section{Co-integration and method}\label{sec:2}

In the process of regression analysis for most non-stationary  time series, the difference
method is usually used to eliminate the non-stationary trend term in the series, so that
the series can be modeled after it is stationary. However, the series themselves after difference
calculation often are limited the scope of the problem discussed and make the reconstructed
model difficult to explain. The co-integration theory has greatly improved the difficulty
of non-stationary series in modeling. The co-integration theory is proposed for integration.
A series with no deterministic component which has a stationary, invertible, ARMA representation
after difference $d$ times, is said to be integrated of order $d$, denoted $Y_t\thicksim I(d)$.
Obviously, for $d=0$, $Y_t$ will be stationary.

The co-integration theory can be understood as that there may be a long-term equilibrium
relationship between several time series with the same order of integration, and one kind
of linear combination of them has a lower order of integration. To formalize these ideas,
the following definition adopted from Engle and Granger is introduced: (i) if all components
of $Y_t$  are $I(d)$; (ii) there exists a vector $\alpha$ ($\neq$ 0) so that
$\alpha^{'} Y_t \thicksim I(d-b)$, $d\geq b\geq 0$. The components of the vector $Y_t$ are said to be
co-integrated of order d, b, denoted $Y_t \thicksim CI(d-b)$, and the vector $\alpha$ is
called the co-integrating vector.

In general, there are two main methods to examine the co-integration, including Engle-Granger
(EG) two-step method and Johansen-Juselius(JJ) multivariate maximum likelihood method
(\citealt{Engle+Granger+1987,Johansen+1995}). The major difference between the above methods is that the EG
two-step method adopts solving linear equation technique, while the JJ test uses the
multivariate equation technique. In this paper, EG method is adopted to assess the null
hypothesis of no co-integration among the time series in $Y_t$. Detailed test steps can be
seen in Ref.(\citealt{Engle+Granger+1987}).

\section{EPT algorithm based on co-integration theory}\label{sec:3}
In pulsar timing, the timing residuals are the differences between the observed
times of arrival (TOAs) and the ones predicted by the timing model, i.e., the difference
between two time scales, AT and PT. Here, AT and PT stand for the atomic time scale and pulsar
time scale, respectively. However, in the practical data processing, the AT recorded the TOAs
should be transferred to Barycentric coordinate time(TCB), and PT is predicted at Solar system
barycenter (SSB) by the pulsar timing model. Hence, for a given pulsar i,
the residuals are denoted as:
\begin{equation}\label{equ1}
Res_i=AT-PT_i \,,
\end{equation}
where $Res_i$ is timing residuals; $AT$ is reference atomic time; $PT_i$ is pulsar time for a
given pulsar i. The EPT (\citealt{Petit+Tavella+1996}) established by multiple pulsars
(i=1, 2 $\cdots$, n) can be defined as:
\begin{equation}\label{equ2}
AT-EPT=\sum\limits_{i=1}^n\omega_i(AT-PT_i),,
\end{equation}
where $\omega_i$ is the relative weight assigned to pulsar i. Because we adopted EG test to
analyze the pulsar data in this paper, hence, assume for two known
pulsars whose timing residuals are $Res_i \thicksim I(1)$, and they are co-integrated,
the co-integrated regression equation of both timing residuals can be expressed as:
\begin{equation}\label{equ3}
Res_1=\alpha+\beta Res_2+\hat{\varepsilon}\,,
\end{equation}
where $\alpha$ and $\beta$ represent regression coefficients; $\hat{\varepsilon}$ is the
model residuals, and ,$\hat{\varepsilon}\thicksim I(0)$.
 According to formula \eqref{equ1}--\eqref{equ3} we obtain
 \begin{equation}\label{equ4}
 \left\{
 \begin{array}{lr}
 \omega_1=\frac{1}{1-\beta}, &\\
 \omega_2=\frac{-\beta}{1-\beta}, &\\
 Res_{ept}=\frac{\hat{\varepsilon}+\alpha}{1-\beta}. &
 \end{array}
 \right.
 \end{equation}
where $Res_{\rm{ept}}$ represents the timing residuals of EPT, it can be regarded as a
transformation from $\hat{\varepsilon}$ by  shift factor $\alpha$
and scale factor $(1-\beta)$, and both $\alpha$ and $(1-\beta)$ are constant coefficients,
which will not affect the order of integration,
so, $Res_{\rm{ept}}\thicksim I(0)$.

\section{Observational data}\label{sec:4}
\subsection{NANOGrav timing observations}\label{sec:4.1}
We used  pulsars timing data from the NANOGrav nine-year data set described in Arzoumanian
et al.(hereafter NG9, \citealt{Arzoumanian+etal+2015b}) for our analysis. NG9 contains 37
MSPs observed at the Green Bank Telescop (GBT) and Arecibo Observatory (AO).  Each telescope
contains two generations of backends, with more recent backends processing up to an order
of magnitude larger bandwidth for improving pulse sensitivity. Polarization calibration and RFI
excision algorithms were applied to the raw data profiles using the PSRCHIVE
(\citealt{Hotan+etal+2004,van+etal+2012}) software package when pulse profiles were folded
and de-dispersed using an initial timing model. After calibration, known RFI signal were
excised, the final pulse profiles used to generate TOAs were fully time averaged with some
frequency averaging to build pulse signal-to-noise ratio (S/N). See NG9 for more detail on
the data processing. Because the purpose of this article is  improve its long-term frequency stability
 of EPT consisting mostly pulsars with significant timing noise, and the stability
 of pulsar timescale is related to the timing span. So, in this paper, the
requirement of selecting pulsars from NG9 includes  that both the sampling time span
is longer than 8 years, and detect obvious evidence for excess at low frequency, or ``red''
timing noise in timing residuals of the pulsars. We selected 7 pulsars that met these
criteria, see the basic parameters of the selected 7 pulsars in Table~\ref{tab1}.

 \begin{table}
\begin{center}
\caption{Basic Parameter of 7 Pulsars}\label{tab1}
\doublerulesep 0.1pt \tabcolsep 4pt 
\begin{tabular}{lccccc}
  \hline\noalign{\smallskip}
 Pulsar name& $P$ (ms) &Number of TOAs &RMS ($\mu$s)& Span (year)&\\
  \hline\noalign{\smallskip}
  J0030+0451 & 4.87 & 2468 & 0.723 & 8.8 \\
  J0613-0200 & 3.06 & 7651 & 0.592 & 8.6 \\
  J1012+5307 & 5.26 & 11995& 1.197 & 9.2 \\
  J1643-1224 & 4.62 & 7119 & 2.057 & 9.0 \\
  B1855+09   & 5.36 & 4071 & 1.339 & 8.9 \\
  J1910+1256 & 4.98 & 2690 & 1.449 & 8.8 \\
  B1937+21   & 1.56 & 9966 & 1.549 & 9.1 \\
 \noalign{\smallskip}\hline
\end{tabular}
\end{center}
\end{table}
\subsection{Data preprocessing}\label{sec:4.2}
NG9 contained all TOAs and timing solutions for 37 pulsars. Each pulsar was observed at each epoch with
at least two receivers. At GBT, the 820 and 1400 MHz bands were used, and at AO, the 430 and 1400 MHz or 1400 and
2300 MHz band were used. We note that frequency-dependent profile shape changes across the entire observing band can be
significant for some sources over the full band(\citealt{Pennucci+etal+2014}), and we wish to maintain homogeneity
of the inferred timing data of our pulsars, we analyze timing residuals with 1400 MHz only.
In addition pulsar B1937+21 just contain data from GBT. We use the unit root test
 with significance level 0.01 to analyze residual datas of 7 pulsars,
the results show that they all seem to be stationary, Res$\thicksim$I(0). It may be due to mostly MSPs in NG9
have higher stability of rotation, the dispersion of
  timing residuals  are dominated by white noise, or red noise is drowned out by white one within
  shorter observation span. Hence, selecting pulsar data with significant red noise and further
  reducing the white noise intensity in the timing residuals in some way that can make the
   processed data meet the condition of non-stationary, Res$\thicksim$I(1), which is equivalent to just retaining the
   red noise component in the original data. Besides, pulsars timing observation are usually irregular
and whose sampling rate are much lower than that atomic clock comparison. Hence, a simple method will be adopted to reduce
the intensity of the white noise and make the two columns of data involved in the co-integration test correspond to each other.

\begin{figure}
\centering
\includegraphics[scale=0.65]{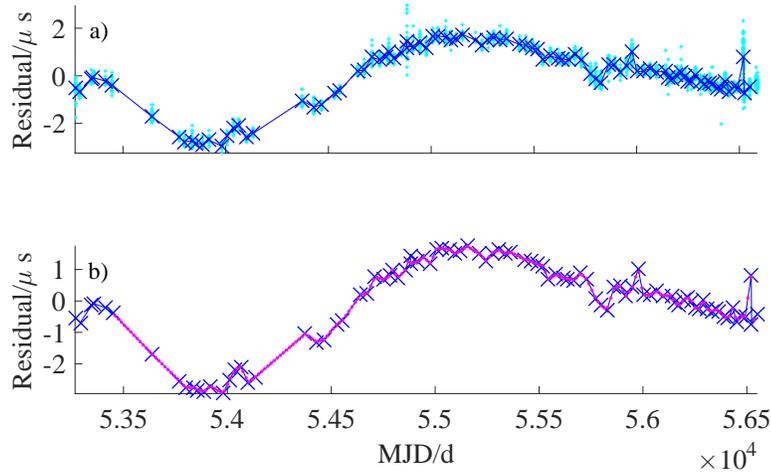}
\caption{The raw timing residuals vs. two-step preprocessing residuals
for pulsar B1937+21 (the cyan and blue represent the raw residuals vs. the
averaged residuals in graph a, and blue and magenta stand for the averaged residuals
 vs. linear interpolation residuals in graph b). } 
\label{fig1}
\end{figure}

For long-term pulsar timing studies it becomes useful to visually inspect timing residuals that have been averaged in order to look for long term trends or biases. The following  details of data preprocessing will be illustrated with one pulsar, i.e., J1937+21: Firstly, we construct daily averaged residuals, each residual value is equal to the average of all raw residuals within one day. This process
 is similar to comparing pulsar time with atomic clock once a day. Subsequently, the data
 are linear interpolated and sampled at intervals of about
 15 days to obtain equally distributed data. The purpose of the above two-step
 is to reduce the  white noise intensity and helpful further analysis whether the low-frequency
 noise in residuals for different pulsars have a co-integration relationship with each other. Other methods
 to  discuss strictly the reduction of white noise intensity will be given in future work.
 The original timing residual distribution vs. two-step preprocessing residual data are
 shown in Figure~\ref{fig1}. Similarly, other six known pulsars timing residuals are also
 regularly processed.


\section{Results and analysis}\label{sec:5}
According to the mathematical model of the co-integration theory in sect.~\ref{sec:3}, it
is necessary to examine integrated order of time series to determine whether there are
co-integrated relationship. In this paper, EG method was adopted to examine timing residuals
of all pulsars after preprocessed data, we found that pulsars B1855+09, B1937+21,
J0030+0451 and J1910+1256 were integrated of order 1, denoted as $I(1)$, and for others were
$I(0)$. These results may be due to the intensity of red noise in residuals of pulsars
J0613-0200, J1012+5307 and J1643-1224 are relatively weak,  after data preprocessing timing residuals  still show
some ``quasi-stationary'' features. In order to search for pulsars with co-integration
relationship by using EG two-step method strictly, we only make further analysis on pulsars
B1855+09, B1937+21, J0030+0451 and J1910+1256.

\begin{figure}
\centering
\includegraphics[scale=0.65]{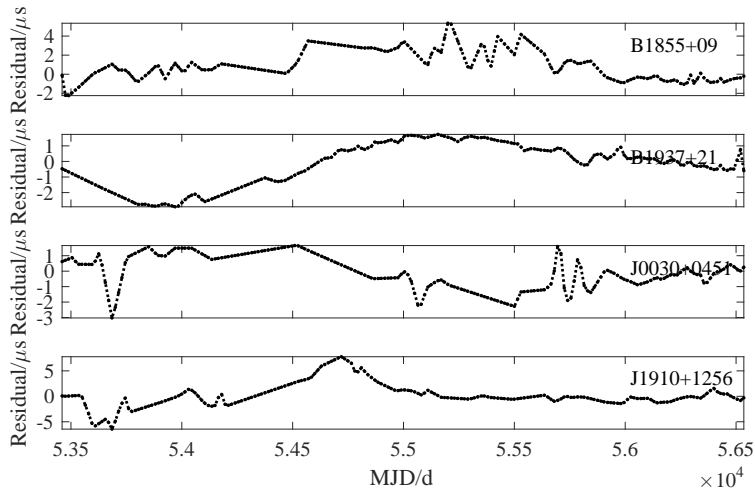}
\caption{The timing residuals for four known pulsars B1855+09, B1937+21, J0030+0451 and
 J1910+1256, respectively.} 
\label{fig2}
\end{figure}

\begin{figure}
\centering
\includegraphics[scale=0.65]{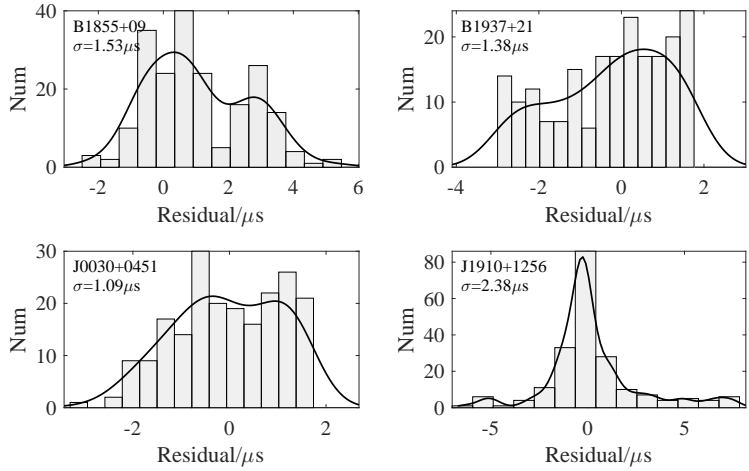}
\caption{The histograms of timing residual distributions for four known pulsars B1855+09,
B1937+21, J0030+045 and J1910+1256, respectively. The solid line is the fitting curve for
each residual distribution. The value of standard deviations ($\sigma$) for four residual
distributions are also given} 
\label{fig3}
\end{figure}

 The timing residuals of pulsars B1855+09, B1937+21, J0030+0451 and J1910+1256 are shown
 in Figure~\ref{fig2}, and the histograms of residuals distribution can be seen in
 Figure~\ref{fig3}, respectively. In Figure~\ref{fig2}, the timing residuals distributions
 of all pulsars show obvious irregular low-frequency trend terms, and histograms of residuals
 distribution are significantly different from the normal distribution in Figure~\ref{fig3}.
 All of these indicate that the timing residuals for four known pulsars have a common feature of instability,
 which is consistent with the case where the integrated order of residuals are denoted as
 $I(1)$. In addition, by comparing the residual distributions, the standard deviations of the
 residuals, all show that there are significant differences for pulsar data each other,
 these differences are not only shown in the trend term of residuals distribution, but
 also in the shape of fitting curve. These are related to the fact that every pulsar data
 is affected by different sources of noise.

\begin{figure}
\centering
\includegraphics[scale=0.65]{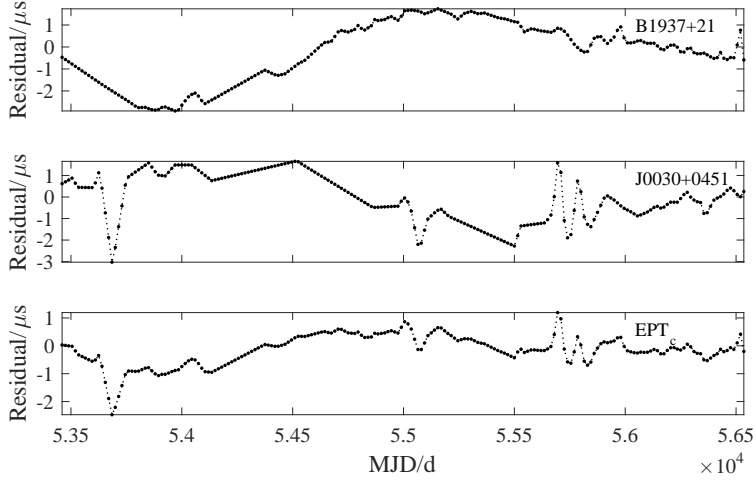}
\caption{The timing residuals for two known pulsars  B1937+21, J0030+0451 and
 for ensemble pulsar timescale $\rm{EPT_c}$, respectively.}
\label{fig4}
\end{figure}

\begin{figure}
\centering
\includegraphics[scale=0.65]{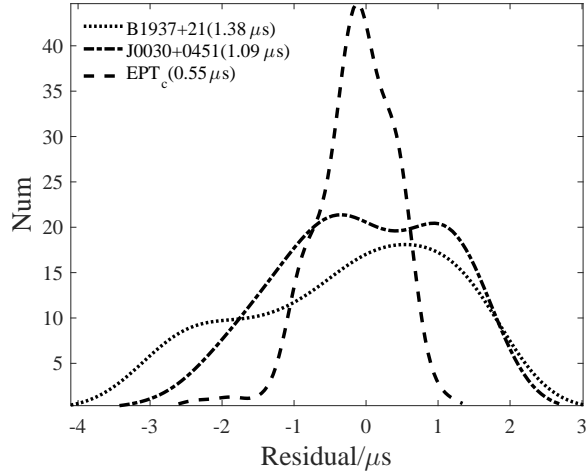}
\caption{The fitting curves of timing residual distributions for two known
pulsars (dotted line for B1937+21 and dash-dot line for J0030+0451) and for ensemble pulsar
timescale (dashed line for $\rm {EPT_c}$), respectively. The value of standard deviation
for three residual distributions are also given in brackets.}
\label{fig5}
\end{figure}

Next, according to formula \eqref{equ3} in sect~\ref{sec:3}, we had further to examine the
linear combination for residuals from two random pulsars (for example pulsars A and B) with denoted
$I(1)$, if $\hat{\varepsilon}$ is integrated of order 0, then pulsars A and B are
co-integrated. We found that only linear combination of pulsars B1937+21 and J0030+0451
had met the condition that the $\hat{\varepsilon}$ is denoted as $I(0)$. So it indicates
that the Pulsars B1937+21 and J0030+0451 are co-integrated. The timing residuals of the
EPT established by the pulsars B1937+21 and J0030+0451 can be obtained according to the formula
\eqref{equ4} in sect~\ref{sec:3} and marked as $\rm{EPT_c}$. In order to obtain the degree
of stability  of residuals, first of all, we compared both residual distributions and residual
histograms for pulsars B1937+21, J0030+0451 and $\rm{EPT_c}$ in Figure~\ref{fig4} and
\ref{fig5}. In Figure~\ref{fig4}, we can see that the amplitude fluctuation of residuals
of pulsars B1937+0451 and J0030+0451 are stronger and have obvious low-frequency freatures,
but the residuals for $\rm{EPT_c}$ are characterized by normalization, simplicity,
significant reduction of non-stationary process, etc. The range of residuals amplitude
variation for pulsars B1937+21, J0030+0451 and $\rm{EPT_c}$ are (-2.91,+1.74) $\mu$s,
(-3.03,+1.66) $\mu$s and (-2.47,+1.19) $\mu$s, respectively. The standard deviation for
$\rm{EPT_c}$ is smallest. In addition, the shape of fitting curve for $\rm{EPT_c}$ in
Figure~\ref{fig5} is also closer to normal distribution than that those of pulsars B1855+09
and J0030+0451. The above contents all indicate that the degree of stability of $\rm{EPT_c}$
has been greatly improved.

\subsection{Variance analysis}\label{sec:5.1}

\begin{figure}
\centering
\includegraphics[scale=0.65]{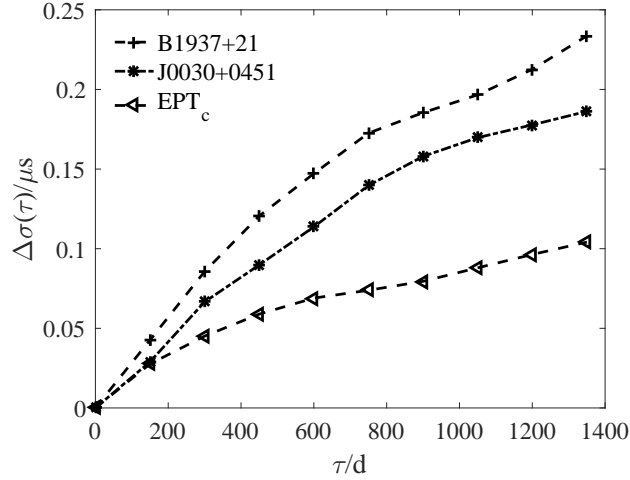}
\caption{The standard deviation increment of timing residuals vs. timing span for two known
pulsars (plus sign for B1937+21 and asterisk for J0030+0451) and for ensemble pulsar timescale
(left-pointing triangle for $\rm{EPT_c}$), respectively.} 
\label{fig6}
\end{figure}

The dispersion of pulsar timing residuals can be divided into white and red noise. White
noise mainly comes from the random errors in the process of timing observation, while red
noise is a kind of signal having strong intensity at lower frequencies, giving it a
power-law spectral density. We can define the dispersion of residual as $\sigma_{\rm RMS}$,
while white noise is denoted $\sigma_{\rm W}$ and red noise is denoted
$\sigma_{\rm TN}$(\citealt{Yang+etal+2014,Gao+etal+2018}). In theory, their relations are
following as:
\begin{equation}\label{equ5}
{\sigma_{\rm RMS}}^2={\sigma_{\rm W}}^2+{\sigma_{\rm TN}}^2 \,.
\end{equation}

where if the ratio of $\sigma_{\rm RMS}$ to $\sigma_{\rm W}$ is close to 1, it indicates
that the dispersion of residual is mainly affected by white noise, and the data is stationary.
If the value of $\sigma_{\rm RMS}/\sigma_{\rm W}$ is much higher than 1, it indicates that
there is a significant red noise component within the timing residuals. Generally, the effect
of red noise on dispersion of residual changes with increasing of observing span, to reflect
this change, the Ref.(\citealt{Gao+etal+2018,Lam+etal+2017}) used variance increment to
show the important contribution of red noise to residual fluctuation. Considering that the
dimension of standard deviation is consistent with the magnitude of data, it is more obvious
when describing data dispersion, and the variance and standard deviation can be easily
converted to each other, as defined by Gao(\citealt{Gao+etal+2018}), the standard deviation
increment is defined as follows:
\begin{equation}\label{equ6}
\Delta\sigma(\tau)=\langle Var(X(t+\tau))^{1/2}-Var(X(t))^{1/2}\rangle \,,
\end{equation}
where $t$ stands for timing span,  $Var(X(t))$ represents the variance of the data in $t$,
and $\tau$ is the increment of timing span. In theory, for data with significant system
fluctuations, the standard deviation increment will change with the increase of $\tau$.
In this paper, we take the pulsar B1937+21 as an example to illustrate how to choose the
values of the parameters $t$ and $\tau$. After data preprocessing, there are 206 points of
residuals for pulsar B1937+21 in 8.4 years, the interval between two points is approximately
15 days. To take $t$ as the interval span of adjacent 10 points, for $\tau$=0, 10, 20,
$\cdots$, add 10 points interval span one by one. Meanwhile, in order to avoid introducing
statistical error, the standard deviation increment in \eqref{equ6} is averaged. The timing
span is divided into two segments at least, so the maximum span of $\tau$ is nearly
half of 8.4 years. Similarly, pulsar J0030+0451 and $\rm{EPT_c}$ are same processed, The
relations between standard deviation increment and timing span for 3 pulsars are shown in
Figure~\ref{fig6}.

As it show that in Figure~\ref{fig6}, the values of $\Delta\sigma(\tau)$ for pulsars B1937+21
and J0030+0451 increase rapidly with the change of $\tau$, while the $\Delta\sigma(\tau)$ for
$\rm{EPT_c}$ changes slowly. This is because the red noise in the residual of pulsars B1937+21
and J0030+0451 are obvious, along the timing span increase, strong red noise becomes an important
factor for the dispersion of the residual of pulsar.  This is consistent with the fact that the
integrated order of two pulsars are 1. It can also be explained that the linear combination
of pulsar residuals which are co-integration is stationary series by $\rm{EPT_c}$.

\subsection{$\sigma_z(\tau)$ methods}\label{sec:5.2}
Using the exceptional rotational stability of millisecond pulsars to generate a time scale
need a reliable statistical measure for studying the physics of pulsar rotation and comparing
pulsar stabilities with those of terrestrial clocks. Clock data are commonly analyzed
using a statistic called $\sigma_y(\tau)$, the square root of the ``Allan variance''
(\citealt{Allan+1966}), which can be computed from second differences of a table of clock
offset measurements. $\sigma_y(\tau)$ is ideally suited for analyzing atomic timescale
which have very small frequency drift rates. However, for most pulsars timing data, the
lowest-order deviations is related to $third$ differences, which is remaining in a pulsar timing
series after the phase, frequency, spin-down rate, and astrometric parameters have been
determined by comparison with terrestrial time, and their effects removed. Following Taylor
(\citealt{Taylor+1991}), the statistic $\sigma_z(\tau)$ defined in terms of third-order
polynomials fitted to sequences of measured time offsets is suggested for studying the
pulsar timing data. Since it is more sensitive to redder noise than other commonly used
measures, and is suited for comparing pulsar stabilities with those of other time scales.
In this paper, we use an improved $\sigma_z(\tau)$  proposed by Matsakis
(\citealt{Matsakis+etal+1997}), which is a good statistic for the analysis of
low-frequency-dominated red noise of pulsar timing residuals. To find $\sigma_z(\tau)$,
divide the data into subsequences, and fit the cubic function to the data in each subsequence
by minimizing the weighted sum of squared differences
\begin{equation}\label{equ7}
R^2=\sum\limits_{i=1}^{N_m}\left[x(t_i)-\frac{X(t_i)}{\sigma_i}\right]^2=min \,,
\end{equation}
Then set
\begin{equation}\label{equ8}
\sigma_z(\tau)=\frac{\tau^2}{2\sqrt{5}}\langle c_3^2\rangle^{1/2}\,.
\end{equation}
where angle brackets denote averaging over the subsequences, weighted by the inverse squares
of the formal errors in $c_3$. The detailed  recipe for the computation of $\sigma_z(\tau)$
can be seen in Ref.(\citealt{Matsakis+etal+1997}).

\begin{figure}
\centering
\includegraphics[scale=0.65]{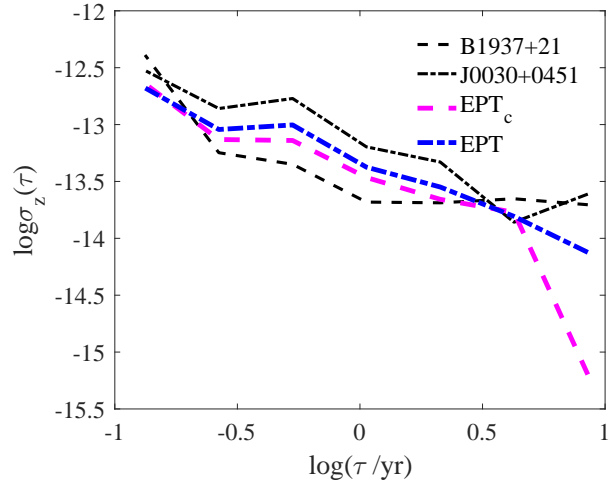}
\caption{Stability of pulsar time scale for two known pulsars (dashed line
for B1937+21 and dash-dot line for J0030+0451) and for ensemble pulsar timescale calculated
by different algorithm (magenta dashed line for $\rm{EPT_c}$ and blue dash-dot line for EPT), respectively.} 
\label{fig7}
\end{figure}

In Figure~\ref{fig7}, we present values of $\sigma_z(\tau)$ for all pulsars B1937+0451,
J0030+0451 and $\rm{EPT_c}$, defined as the weighted root-mean-square of the coefficients
of the cubic terms fitted over intervals of length $\tau$. For comparison, another
EPT calculated by traditional classical weighted average algorithm is given in Figure~\ref{fig7}, the weights $\omega_i$
are inversely proportional to the variance of two pulsars 1937+21 and J0030+0451, respectively. Because $\sigma_z^2(\tau)$ can
be easy to describe by a power law, if the white noise is dominate in the time series, the
slope of the log-log graph is close to -1.5, for all four time series show that at least
up to intervals $\tau$ of several years, as expected for residuals dominated by uncorrelated
measurement errors. On the contrary, when the red noise is dominant, the tail of the curve
will gradually become an upward trend, which is represented as the influence of low-frequency
noise on the frequency stability. For both pulsars B1937+21 and J0030+0451, the curves show
a tail upward trend, while the curves of both $\rm{EPT_c}$ and EPT show a downward trend as a whole,  it
indicates that the timing residuals for two pulsars B1937+21 and J0030+0451 are dominated by
low-frequency noise believed to be intrinsic to the pulsars.  It is note that the
stability of pulsar time scale for pulsar J0030+0451 at $\log\tau\thicksim$0.6 is more stable
than $\rm{EPT_c}$. This anomaly should be induced by the increasing errors of $\sigma_z$ for
larger intervals of length $\tau$ because of the decreasing number of sequences .

The long-term frequency stability level plays an important role in the study of pulsar time
scales, Figure~\ref{fig7} shows that the value of $\sigma_z(\tau)$ for two pulsars B1937+0451 and
J0030+0451 and for ensemble pulsar timescale calculated by different algorithm $\rm{EPT_c}$ and
EPT on the span 8.4 years are 10$^{-13.70}$, 10$^{-13.61}$,
10$^{-15.20}$ and 10$^{14.12}$, respectively. The long-term frequency stability of $\rm{EPT_c}$ is nearly
one order of magnitude higher than those of the other two pulsars B1937+0451 and
J0030+0451. In addition, one can note that the stability of $\rm{EPT_c}$ is better than EPT as a whole in
Figure~\ref{fig7}. The above analysis indicates that
the long-term frequency stability level of pulsar can be significantly improved within
a limited observation span when combining the pulsars data with co-integration relation
to establish EPT. One thing to keep in mind is that these stabilities of pulsars in
Figure~\ref{fig7} are not further compared with both other pulsars data with weak red noise
and  atomic timescale at different stations, this is because there are many factors influencing
the stabilities of different timescale. In this paper, we only show that the method to establish
EPT based on combining the pulsars data with the co-integration relation is reliable and
feasible, further research will be carried out in the follow-up work.

\section{Discussion and Conclusions}\label{sec:6}
It should be noted that it is necessary to satisfy a strict constraint condition to establish
EPT based on the co-integration theory: the timing residuals of pulsars are co-integration.
The combination of timing residuals with co-integration relation can only reduce the number
of integrated order, and in this way it can obviously reduce the timing noise intensity of
EPT and improve its long-term frequency stability. According to this constraint condition,
it means that the following problems may be encountered in practical application: (1) In
the short span, the red noise in residuals of millisecond pulsars are not dominant or can
not be measured, residuals data show ``quasi-stationary'' characteristics , so it will
limit the methods in application pulsars data. (2) At present, the low-frequency timing noise
of most pulsars data are irregular, which means that the co-integration relation between
pulsars may be damaged by the increase or decrease of the data span,  it may lead to segmented
co-integration. In addition, we only use the EG(\citealt{Engle+Granger+1987}) two-step test
to discuss and establish EPT algorithm based on two pulsars in this paper, if the data of
multiple pulsars are regarded as multivariate variables, and the co-integration relation
between them is tested by using JJ(\citealt{Johansen+1995}) methods , the application of
co-integration theory in the algorithm of EPT can be further extended.

Based on the co-integration theory, an algorithm to establish EPT by using pulsars data with
significant timing noise is proposed in this paper, this algorithm can successfully reduce
several timing noises and improve the long-term stability of pulsar timescale.
Compared to the optimal weighting method (\citealt{Rodin+2008}) and
the global fitting method (\citealt{Hobbs+etal+2012}), our co-integration method
is similar to  the traditional classical weighted average method but with a new way
of choosing weights. However, different
from the traditional classical weighted average algorithm , this algorithm can effectively suppress
some noise sources if there is a co-integration relationship between different pulsar data,
and provides the chances of the pulsar with significant red noises to attend the establishment
of ensemble pulsar timescale.

\begin{acknowledgements}
This work was funded by the National Natural Science Foundation of China No. 11373028, U1531112, 91736207,
11873050, 11873049, U1831130,  The A Project of the Young Scholar of the ``West Light''
of the Chinese Academy of Sciences No. XAB2015A06, and Cultivation Found of Xi'an University of Science and
Technology No. 201707.
\end{acknowledgements}

\label{lastpage}


\begin{thebibliography}{99}

    \bibitem[Allan(1966)]{Allan+1966} Allan D. W. 1966, In Proceedings of the IEEE, 54, 211

     \bibitem[Arzoumanian et al.(2015a)]{Arzoumanian+etal+2015a} Arzoumanian Z., Brazier A., Burke-Spolaor S., et al. 2015a, \apj, 810, 150

     \bibitem[Arzoumanian et al.(2015b)]{Arzoumanian+etal+2015b} Arzoumanian Z., Brazier A., Burke-Spolaor S., et al. 2015b, \apj, 813, 65

     \bibitem[Coles et al.(2011)]{Coles+etal+2011} Coles W., Hobbs G., Champion D. J., Manchester R. N.,Verbiest J. P. W. 2011, MNRAS, 418, 561

     \bibitem[Engle \& Granger(1987)]{Engle+Granger+1987} Engle Robert F., Granger C. W. J. 1987, Econometrica,  55, 251

    \bibitem[Gao et al.(2018)]{Gao+etal+2018} Gao F., Gao Y.-P., Tong M.-L., Yang T.-G., Zhao C.-S. 2018, (in Chinese)Sci Sin-Phys Mech Astron,  48: 059501

     \bibitem[Hobbs et al.(2010)]{Hobbs+etal+2010} Hobbs G., Coles W., Manchester R., Chen D. 2010, preprint (arXiv: 1011.5285)

     \bibitem[Hobbs et al.(2012)]{Hobbs+etal+2012} Hobbs G., Coles W., Manchester R. N., Keith M. J., et al. 2012, MNRAS, 427, 2780

     \bibitem[Hotan et al.(2004)]{Hotan+etal+2004}Hotan A., van Straten W., Manchester R. N. 2004, PASA, 21, 302

     \bibitem[Johansen(1995)]{Johansen+1995} Johansen S. 1995, (Oxford: Oxford University Press)

    \bibitem[Lam et al.(2017)]{Lam+etal+2017} Lam M. T., Cordes J. M., Chatterjee S., et al. 2017, \apj , 834, 35

    \bibitem[Matsakis et al.(1997)]{Matsakis+etal+1997} Matsakis D. N., Taylor J. H., Eubanks T. M. 1997, A\&A,  326, 924

    \bibitem[Petit et al.(1993)]{Petit+etal+1993} Petit G., Thomas C., Tavella P. 1993, The 24th
    Annual Precise Time and Time Interval Applications and Planning Meeting,(Sevres CEDEX: NTRS) 73

    \bibitem[Petit \& Tavella(1996)]{Petit+Tavella+1996} Petit G., Tavella P. 1996, A\&A, 308, 290
    \bibitem[Pennucci et al.(2014)]{Pennucci+etal+2014}Pennucci, T. T., Demorest, P. B., \& Ransom, S. M. 2014, \apj,790, 93

    \bibitem[Rodin(2008)]{Rodin+2008} Rodin A. E. 2008,  Monthly Notes of the Astronomical Society of the South Africa, 387, 1583

    \bibitem[Splaver(2004)]{Splaver+2004} Splaver E. M. 2004, Ph. D Dissertation. Dept. of Physics, Princeton University

    \bibitem[Taylor(1991)]{Taylor+1991} Taylor J. H. 1991,  In Proceedings of the IEEE,79, 1054

     \bibitem[Tong et al.(2017)]{Tong+etal+2017} Tong M.-L., Yang T.-G., Zhao C.-S., Gao Y.-P. 2017,  (in Chinese)Sci
Sin-Phys Mech Astron,  44: 099503

     \bibitem[van Straten et al.(2012)]{van+etal+2012}van Straten W., Demorest P., Oslowski S. 2012, AR\&T, 9, 237

     \bibitem[Verbiest et al.(2009)]{Verbiest+etal+2009} Verbiest J. P. W., Bailes M., Coles W. A., et al. 2009, MNRAS, 400, 951

      \bibitem[Will(2014)]{Will+2014} Will C. M., 2014, LRR, 17, 4

      \bibitem[Yang et al.(2014)]{Yang+etal+2014} Yang T.-G., Tong M.-L., Gao Y.-P. 2014, (in Chinese)J Time Freq, 37, 80

       \bibitem[Zhong \& Yang(2007)]{Zhong+Yang+2007} Zhong C.-X., Yang T.-G. 2007, (in Chinese)Acta Phys. Sin. Vol. 56, No. 10

    \bibitem[Zhu et al.(2015)]{Zhu+etal+2015} Zhu W.-W., Stairs I. H., Demorest P. B., et al. 2015, \apj, 809, 41

\end{thebibliography}
\end{document}